%% file: main.tex
\documentclass[sigconf]{acmart}

\usepackage{booktabs} 
\usepackage{multirow}
\usepackage{subfigure}
\usepackage{verbatim}
\usepackage{CJKutf8}
\usepackage{flushend}

\usepackage{makecell}
\usepackage[utf8]{inputenc}

\copyrightyear{2022}
\acmYear{2022}
\setcopyright{acmcopyright}
\acmPrice{15.00}

\begin{document}

\title{End-to-end Learnable Diversity-aware News Recommendation}

\fancyhead{}


\author{Chuhan Wu$^1$, Fangzhao Wu$^2$, Tao Qi$^1$, Yongfeng Huang$^1$}

\affiliation{%
  \institution{$^1$Department of Electronic Engineering, Tsinghua University, Beijing 100084 \\ $^2$Microsoft Research Asia, Beijing 100080, China}
} 
\email{{wuchuhan15,wufangzhao,taoqi.qt}@gmail.com,yfhuang@tsinghua.edu.cn}






\begin{abstract}
Diversity is an important factor in providing high-quality personalized news recommendations.
However, most existing news recommendation methods only aim to optimize recommendation accuracy while ignoring diversity.
Reranking is a widely used post-processing technique to promote the diversity of top recommendation results.
However, the recommendation model is not perfect and errors may be propagated and amplified in a cascaded recommendation algorithm.
In addition, the recommendation model itself is not diversity-aware, making it difficult to achieve a good tradeoff between recommendation accuracy and diversity.
In this paper, we propose a news recommendation approach named \textit{LeaDivRec}, which is a fully learnable  model that can generate diversity-aware news recommendations in an end-to-end manner.
Different from existing news recommendation methods that are usually based on point- or pair-wise ranking, in \textit{LeaDivRec} we propose a more effective list-wise news recommendation model.
More specifically, we propose a permutation Transformer to consider the relatedness between candidate news and meanwhile can learn different representations for similar candidate news to help improve recommendation diversity.
We also propose an effective list-wise training method to learn accurate ranking models.
In addition, we propose a diversity-aware regularization method to further encourage the model to make controllable diversity-aware recommendations.
Extensive experiments on two real-world datasets validate the effectiveness of our approach in balancing recommendation accuracy and diversity.

\end{abstract}

%
%



\maketitle

\input{data/introduction.tex}

\input{data/relatedwork.tex}

\input{data/method.tex}

\input{data/experiment.tex}

\input{data/conclusion.tex}

\bibliographystyle{ACM-Reference-Format}
\bibliography{main}

\end{document}

%% file: data/introduction.tex
\section{Introduction}

Online news services such as the Google News\footnote{https://news.google.com/} website and the ``News and interests'' app on Windows have attracted a large number of users to read digital news~\cite{das2007google,wu2020mind}.
However, the explosion of online news information leads to heavy information overload of users~\cite{wu2019npa}.
News recommendation techniques can help provide personalized news services for users to improve their online news reading experience~\cite{okura2017embedding}.

Most existing news recommendation methods rank candidate news according to their relevance to user interest~\cite{okura2017embedding,wang2018dkn,wu2019nrms,wang2020fine}.
For example, 
Okura et al.~\shortcite{okura2017embedding} proposed to use denoising autoencoders to learn news representations and use a GRU network to learn user representations from clicked news.
They ranked candidate news based on the relevance score computed by the relevance between candidate news and user embeddings.
Wu et al.~\shortcite{wu2019nrms} proposed to learn representations of news and users with multi-head self-attention networks, and rank candidate news based on the relevance between news and user representations.
Wang et al.~\cite{wang2020fine} proposed a fine-grained interest matching approach that models the relevance between words in candidate news and clicked news for computing ranking scores.
However, these methods mainly aim to optimize recommendation accuracy while neglecting the diversity of recommendation results, which is also critical for user experience.
In addition, they compute ranking scores for different candidate news independently and ignore the relatedness between the news in the candidate list, which may be suboptimal for making accurate and diverse recommendations.

Reranking is a widely used post-processing scheme for promoting the diversity of top recommendation results~\cite{carbonell1998use}.
For example, Determinantal Point Processes (DPP)~\cite{chen2018fast} is a popular technique for reranking, which can control the tradeoff between recommendation relevance and diversity by optimizing the determinant of a kernel matrix.
It has been served as the core component of many reranking-based diversity-aware recommendation methods~\cite{wilhelm2018practical,wu2019pd}.
However, reranking-based methods can only modify the off-the-shelf recommendation list based on the relevance score predicted by the recommendation model, where the errors of the model may be propagated and even amplified in the cascaded algorithm.
In addition, the recommendation model is usually not diversity-aware, making it difficult for the subsequent reranking process to achieve a good tradeoff between recommendation accuracy and diversity.

In this paper, we propose a \underline{\textbf{lea}}rnable \underline{\textbf{div}}ersity-aware news \underline{\textbf{rec}}ommendation approach  (\textit{LeaDivRec}), which can make accurate and diverse news recommendations in an end-to-end manner.
Instead of measuring the relevance between user interest and each candidate news independently, we propose a list-wise ranking model that can fully consider the relatedness between candidate news to be ranked.
More specifically, we propose a permutation Transformer with different input permutation orders in different attention heads to capture the relations between candidate news, and meanwhile enables learning diverse representations for similar news to help make diverse recommendations.
In addition, we propose an effective list-wise model training method to learn an accurate news ranking model.
Besides, to help the model satisfy the requirements under different diversity intensities, we propose a diversity-aware regularization method to encourage selecting news with diverse information in the top recommendation results, which is controlled by a loss coefficient in model training.
Extensive experiments on two real-world news recommendation datasets validate that \textit{LeaDivRec} can achieve better tradeoffs between recommendation accuracy and diversity than many baseline methods.

The contributions of this paper are listed as follows:
\begin{itemize}
    \item To our best knowledge, this is the first end-to-end diversity-aware news recommendation method that is free from post-processing techniques like reranking.
    \item We propose a list-wise news ranking model with a novel permutation Transformer architecture to capture the relatedness between different candidate news.
    \item We propose a simple yet effective list-wise model training method with diversity-aware regularization to learn both accurate and diversity-aware recommendation models.
    \item We conduct extensive experiments on two real-world datasets and the results show our approach can achieve good tradeoff between recommendation accuracy and diversity.
\end{itemize}

%% file: data/relatedwork.tex
\section{Related Work}\label{sec:RelatedWork}

\subsection{News Recommendation}

News recommendation has been extensively studied in recent years~\cite{wu2021personalized}.
A core problem in news recommendation is matching candidate news with user interest based on their relevance.
For example, Okura et al.~\shortcite{okura2017embedding} proposed to first use denoising autoencoders to learn news representations, then use a GRU network to learn user representations from clicked news representations, and finally evaluate the relevance between candidate news and user interest based on the inner product between their representations.
Wang et al.~\shortcite{wang2018dkn} proposed to use a knowledge-aware CNN model to learn news representations and use a candidate-aware network to learn user representations with respect to different candidate news.
The final news ranking is based on the relevance score computed from the concatenation of news and user representations.
Wu et al.~\shortcite{wu2019nrms} proposed to use multi-head self-attention networks to learn both news and user representations, and compute relevance scores between them via inner product.
Wang et al.~\shortcite{wang2020fine} proposed a fine-grained interest matching method that uses a 3-D CNN model to compute the click score by capturing the interactions between words in candidate news and clicked news.
Qi et al.~\shortcite{qi2021kim} proposed a knowledge-aware interactive matching method that can model the relatedness between candidate news and clicked news in terms of their semantic and knowledge information in user interest matching.
These methods mainly focus on optimizing the accuracy of recommendation by recommending news that better match users' personal interest.
However, the diversity of recommendation results, which  usually has a huge impact on long-term user experience and engagement, is not considered by these methods.

There are only a few news recommendation methods that explicitly incorporate diversity-aware mechanisms~\cite{li2011scene}.
For example, Li et al.~\shortcite{li2011scene} proposed to rerank candidate news in the same topic category based on their popularity, which can promote diversity because popular news are usually diverse.
Zheng et al.~\shortcite{zheng2018drn} proposed a deep  reinforcement learning based method that uses an exploration network to generate diverse recommendation results and combines them with the recommendation lists generated by the Q-network.
However, in these methods the personalized ranking model is not diversity-aware and its errors may be propagated through the cascaded systems, which is not beneficial for balancing recommendation accuracy and diversity effectively.
Wu et al.~\shortcite{wu2020sentirec} proposed a sentiment diversity-aware news recommendation method that uses a sentiment regularization loss to encourage recommending news with sentiment diverse from clicked news.
However, this method aims to optimize temporal diversity rather than the diversity within a recommendation list.
Different from these existing methods, our approach is an end-to-end approach to promote the diversity of news recommendation results and can meanwhile control the intensity of diversity.

\subsection{Diversity-aware Recommendation}

Diversity-aware recommendation is a widely explored in the recommender system community and there are many genres of technologies such as contextual bandit~\cite{li2010contextual}, regularization~\cite{qin2013promoting,cheng2017learning}, and reranking~\cite{chen2018fast,liu2020diversified}.
Contextual bandit based methods aim to decide whether to explore recommending new items (i.e., exploration) or stick on items  that are relevant to user preferences (i.e., exploitation) in the next recommendation round according to users' feedback.
However, they are usually used in interactive online recommender systems, while most news recommendation models are first trained on offline user logs~\cite{sanz2019simple}.
Regularization-based methods usually jointly train the model with both recommendation loss and diversity-related loss functions.
For example, Qin et al.~\shortcite{qin2013promoting} proposed an entropy regularization method based on  probabilistic matrix factorization (PMF)~\cite{mnih2008probabilistic} by encouraging recommending items with high uncertainties of rating distributions given observations of rated items.
Cheng et al.~\shortcite{cheng2017learning} proposed to first use heuristic rules to find a set of relevant and diverse items as regularization targets and then train a diversified collaborative filtering model.
However, these methods are not suitable for the news recommendation scenario due to the heavy item cold-start problem brought by the quick vanishment of old news and the emergence of newly published news.

\begin{figure*}[!t]
  \centering
    \includegraphics[width=0.99\linewidth]{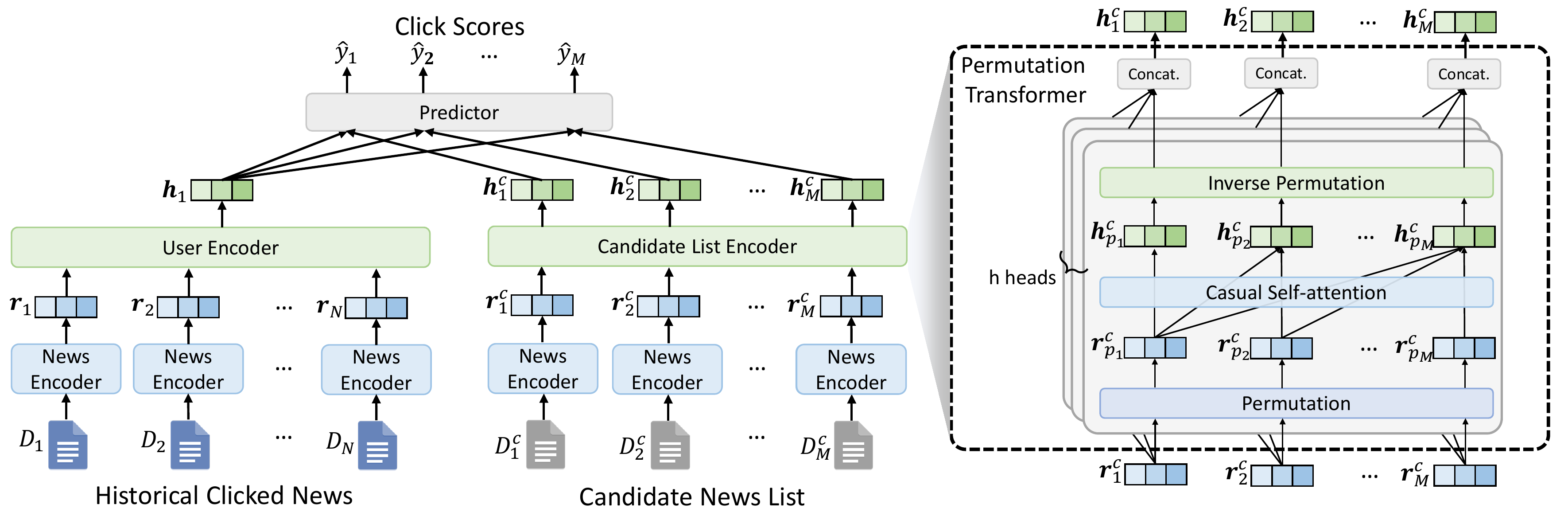}
  \caption{The overall framework of \textit{LeaDivRec}.}
  \label{fig.model}
\end{figure*}

In recent years, determinantal point process (DPP)~\cite{kulesza2012determinantal} based reranking  has become a common fashion in diversity-aware recommendation~\cite{liu2020diversified}.
For example, Wilhelm et al.~\shortcite{wilhelm2018practical} explored using DPP to improve the diversity of Youtube video recommendation based on embedding similarities and the relevance scores predicted by a point-wise ranker. 
Since the inference computational cost of DPP is very high, Chen et al.~\cite{chen2018fast} proposed a fast greedy method for DPP inference to promote recommendation diversity.
This method is also capable of controlling the tradeoff between accuracy and diversity by introducing an additional coefficient.
Wu et al.~\shortcite{wu2019pd} proposed to use an adversarial learning framework to enhance DPP-refined recommendation.
They first used DPP to generate diversified recommendations, then used a discriminator to distinguish between the recommendation list given by a generator and randomly selected items from different categories that a user has interactions with.
However, in these methods the core diversification module is DPP while the recommendation models are usually not diversity-aware.
In addition, the errors encoded by the recommendation models may also influence the subsequent reranking, which may be suboptimal in achieving a good accuracy-diversity tradeoff.
Different from these methods, our approach is an end-to-end framework that can jointly optimize recommendation accuracy and diversity, which can better balance them in a unified way.

%% file: data/method.tex
\section{Learnable Diversity-aware News Recommendation}\label{sec:Model}

In this section, we introduce our proposed learnable diversity-aware news recommendation method named \textit{LeaDivRec}. 
We first present a formal definition of the problem studied in this paper, then introduce the details of our approach, and finally present some analysis on the computational complexity of our approach.

\subsection{Problem Formulation}

Assume a user $u$ has $N$ historical clicked news, which are denoted as $[D_1, D_2, ..., D_N]$.
The candidate news list to be ranked is denoted as $[D^c_1, D^c_2, ..., D^c_M]$, where $M$ is the list length.
Each news is represented by its texts, such as news titles.
The model aims to predict the click scores for the candidate news list, which are denoted as $[\hat{y}_1, \hat{y}_2, ..., \hat{y}_M]$.
In this paper, we use the semantic distance between news to measure diversity.
We denote the pair-wise semantic similarity matrix of  news within a candidate list as $\mathbf{S}$.
The goal of the recommendation model is to rank the candidate news according to their relevance to user interest while keeping diversity among top ranked news based on the metrics derived from $\mathbf{S}$.

\subsection{Diversity-aware List-wise Ranking Model}

Next, we introduce the diversity-aware list-wise ranking model in \textit{LeaDivRec}, which is shown in Fig.~\ref{fig.model}.
The core of this model includes a user model that aims to learn user interest embedding from the sequence of historical clicked news, a candidate list model that aims to model the content of candidate news and their relatedness, and a click predictor that computes the click scores for each news in the candidate news list.
Their details are introduced as follows.

The user model used in our approach is a variant of the user model in NRMS~\cite{wu2019nrms}.
More specifically, we first use a shared news encoder with a Transformer~\cite{vaswani2017attention} layer to capture contextual information in news texts, an attention pooling layer to learn news embeddings, and a dense layer with ReLU activation function to further learn hidden representations of news.
Following many prior works~\cite{wang2018dkn,wu2019npa}, we use news titles to model news.
We denote the hidden representations of clicked news as $[\mathbf{r}_1, \mathbf{r}_2, ..., \mathbf{r}_N]$.
We then use a user encoder to learn a user interest embedding $\mathbf{u}$ from clicked news embeddings.
We also use a Transformer to capture the contexts of click behaviors, and use an attention pooling network to learn user embeddings.
Note that both Transformers in the news and user encoder do not use layer normalization techniques because we find the performance is quite unsatisfactory when they are used.

The candidate list model first uses news encoders to learn content-based representations of news in the candidate list, which are denoted as $\mathbf{R}^c=[\mathbf{r}^c_1, \mathbf{r}^c_2, ..., \mathbf{r}^c_M]$.
Since the candidate news within the same list to be ranked may have some relatedness, we propose to use a candidate list encoder to learn hidden candidate representations.
However, the design of the candidate list encoder has two main challenges.
First, the candidate news list is order-agnostic, and it is not appropriate to use order-sensitive models such as CNN, RNN and Transformer to process the candidate news list.
Thus, the model should be relatively stable to the orders of input candidate news. 
Second, to achieve diversified recommendations, the representations of similar news need to be diverse.
In an extreme case, if there are two identical news in the candidate list, the model needs to learn different representations for them so that their ranking positions can be different.
Thus, order-agnostic models such as self-attention and MLP cannot handle this challenge.
Third, the model needs to support processing relatively long lists with tens or even hundreds of candidates in an end-to-end way.
To solve the challenges mentioned above, inspired by the permutation-invariant model learning mechanism studied in~\cite{murphy2019janossy}, we propose a permutation Transformer that can learn different representations for very similar news and meanwhile keep low order sensitivity.
More specifically, in each attention head we randomly permute the candidate news representations.
To ensure effective matrix operation, we use a permutation matrix $\mathbf{P}$ to shuffle the orders of candidate news representations.
We denote the permuted representation sequence in this head as  $\mathbf{R}^c_p=[\mathbf{r}^c_{p_1}, \mathbf{r}^c_{p_2}, ..., \mathbf{r}^c_{p_M}]= \mathbf{P}\mathbf{R}^c$.
We use a casual self-attention network~\cite{shen2018disan} to capture the relatedness between different news.
Different from the standard self-attention network where each position can attend to all positions in the input sequence, casual self-attention only allows attending to past positions.
We denote the hidden candidate news representations it learns as $\mathbf{H}^c_p=[\mathbf{h}^c_{p_1}, \mathbf{h}^c_{p_2}, ..., \mathbf{h}^c_{p_M}]$.
Finally, we use the inverse permutation matrix $\mathbf{P}^\top$ to recover the orders of candidate news representations, which is achieved by $\mathbf{P}^\top\mathbf{H}^c_p$.
We concatenate the hidden candidate news representations learned by different heads to form the unified candidate news embeddings.
We add residual connections between the output and input of the candidate list encoder, and the final representation sequence is denoted as $[\mathbf{h}^c_{1}, \mathbf{h}^c_{2}, ..., \mathbf{h}^c_{M}]$.
Since in each attention head the casual self-attention is order sensitive, even identical news can generate different representations.
At the same time, since there are multiple attention heads to process permuted sequences independently, the entire permutation Transformer is order-agnostic in the sense of expectation.
Moreover, self-attention is efficient on GPU when the candidate list is not extremely long (e.g., hundreds of candidates).
Thus, our proposed permutation Transformer can well address the three challenges and learn diversified candidate news embeddings.

On the basis of the user interest embedding and the diversified candidate news embeddings, the click predictor predicts the click scores of candidate news based on their relevance to user interests.
Following many existing methods~\cite{okura2017embedding,wu2019nrms}, we use inner product as the prediction function.
For the $i$-th candidate news, its click score $\hat{y}_i$ is computed by $\hat{y}_i=\mathbf{u}^\top \mathbf{h}^c_i$.
Candidate news will be ranked by the click scores, where news with higher click scores will be assigned higher ranks.

\subsection{List-wise Model Training}

We then introduce how to train our proposed list-wise ranking model.
In many existing news recommendation methods~\cite{wu2019npa,wu2019nrms}, negative sampling methods are used to construct labeled training samples, where each clicked news is associated with several non-clicked news.
However, this method is not suitable for training our list-wise ranking model, because users may have multiple clicks on the candidate list.
In addition, it is not suitable to simply model the training task as a binary classification problem at each position because of the highly imbalanced class distribution and the dependency among different candidate news.
To solve this issue, we propose a list-wise contrastive training method.
We denote the predicted click scores of clicked samples and non-clicked samples in the candidate news list as $[\hat{y}_{p_1}, \hat{y}_{p_2}, ..., \hat{y}_{p_P}]$ and  $[\hat{y}_{n_1}, \hat{y}_{n_2}, ..., \hat{y}_{n_Q}]$, respectively.
For each clicked news, we pack it with other non-clicked news and normalize their click scores via softmax.
The loss function $L_{rec}$ is the total summation of the negative log-likelihood of the normalized click scores of all clicked news, which can be formulated as follows:
\begin{equation}
    \mathcal{L}_{rec}=-\sum_{i=1}^P\log(\frac{\exp(\hat{y}_{p_i})}{\exp(\hat{y}_{p_i})+\sum_{j=1}^Q\exp(\hat{y}_{n_j})}).
\end{equation}
Note that in this formula, if a candidate list has more clicked news, this training sample will gain a higher loss, which means that this sample is more important.
In addition, the relatedness between candidate news can also be taken into consideration.
However, it is insufficient to train the model with click signals only to make diversified recommendations.
Thus, we propose an additional diversity-aware regularization method to encourage the model to generate diverse recommendation results.
The diversity-aware regularization loss function is formulated as follows:
\begin{equation}
    \mathcal{L}_{div}=\sum_{i=1}^M\sum_{j=1}^M \hat{y}_i \hat{y}_j s_{i,j},
\end{equation}
where $s_{i,j}$ denotes the semantic similarity between the $i$-th and $j$-th candidate news.
In this formula, if both the $i$-th and $j$-th candidate news have high click scores, the regularization loss will be large if they are similar.
In addition, if one of two similar news has a low click score, the regularization loss will be small because they are ranked at very different positions or both at low positions.
In this way, the model can be encouraged to assign similar candidate news to different ranks.
We jointly train the ranking model in both the list-wise contrastive training loss and the diversity-aware regularization loss.
The unified loss function $\mathcal{L}$ is a weighted summation of both loss functions, which is formulated as follows:
\begin{equation}
    \mathcal{L}=\mathcal{L}_{rec}+\lambda\mathcal{L}_{div},
\end{equation}
where $\lambda$ is a hyperparameter that controls the intensity of diversity impact on the recommendation results.
By optimizing the overall loss function in model training in an end-to-end manner, the recommendation model can generate both accurate and diverse recommendation results.

\subsection{Complexity Analysis}

Finally, we discuss the computational complexity of \textit{LeaDivRec}.
Since the user model and click predictor are common modules used by many methods, we only discuss the complexity of the candidate list model and model training.
In the permutation Transformer each head uses a casual self-attention to process the candidate news embedding sequence.
Thus, its computational complexity is $O(N^2d)$, where $d$ is the hidden dimension.
The computational cost of obtaining the semantic similarity matrix among different candidate news is also $O(N^2d)$.
In the test phase, the candidate news list to be ranked is usually not very long (with tens or hundreds of news), and the user embedding, candidate news embeddings and semantic similarity matrix can all be computed in advance and cached.
Thus, the computational cost of our approach is acceptable.

%% file: data/experiment.tex
\section{Experiments}\label{sec:Experiments}

\subsection{Datasets and Experimental Settings}

We use two real-world datasets to conduct our experiments.
The first dataset is  MIND~\cite{wu2020mind}\footnote{https://msnews.github.io/.}, which is a large-scale English benchmark dataset for news recommendation.
It contains 1 million users' news click logs on Microsoft News during a period of six weeks.
The second dataset is a proprietary dataset collected by ourselves from a commercial news platform.
We denote this dataset as \textit{PrivateNews}.
It is constructed from 1 million news impression logs from Oct. 17, 2020 to Jan. 29, 2021.
The anonymous users IDs are delinked from the original production IDs via salted hash to protect user privacy.
In both datasets, the impressions in the last week are used as the test sets, the impressions on the last day before the test week are used for validation, and the rest are used for training.
The statistics of \textit{MIND} and \textit{PrivateNews} are shown in Table~\ref{dataset}.

\begin{table}[h]
\centering
\caption{Detailed statistics of the two datasets.}\label{dataset}
\begin{tabular}{lcc}
\Xhline{1.5pt}
                   & \textit{MIND} & \textit{\textit{PrivateNews}} \\ \hline
\# Users           & 1,000,000     & 1,322,973             \\
\# News            & 161,013       & 4,378,487             \\
\# Impressions     & 15,777,377    & 1,000,000             \\
\# Click Behaviors & 24,155,470    & 41,976,699             \\
\# Avg. Title Len. & 11.52         & 12.62           \\ \Xhline{1.5pt}
\end{tabular}
\end{table}

\begin{figure}[!t]
	\centering
	\subfigure[\textit{MIND}.]{
	\includegraphics[width=0.35\textwidth]{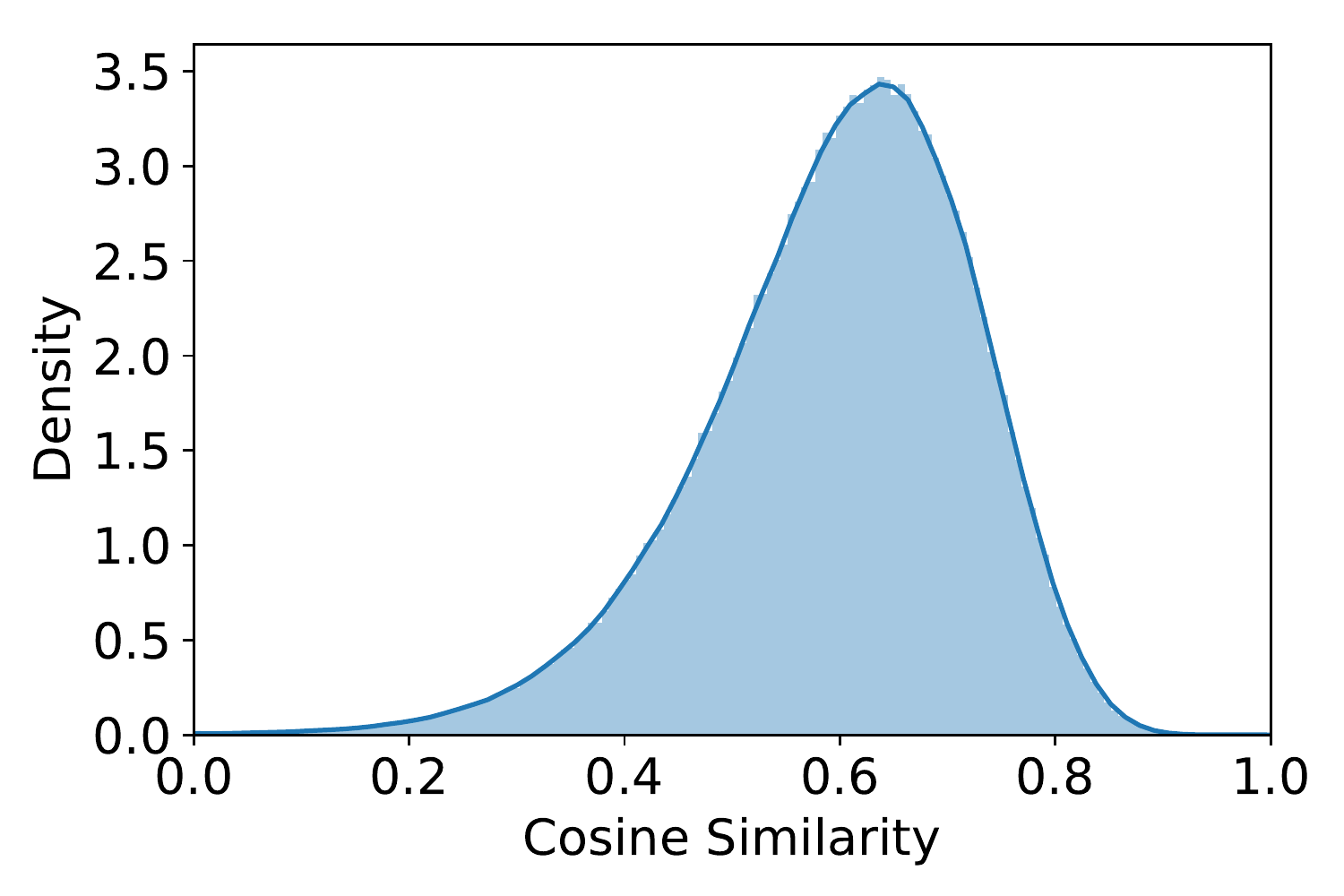}
	}
		\subfigure[\textit{PrivateNews}.]{
	\includegraphics[width=0.35\textwidth]{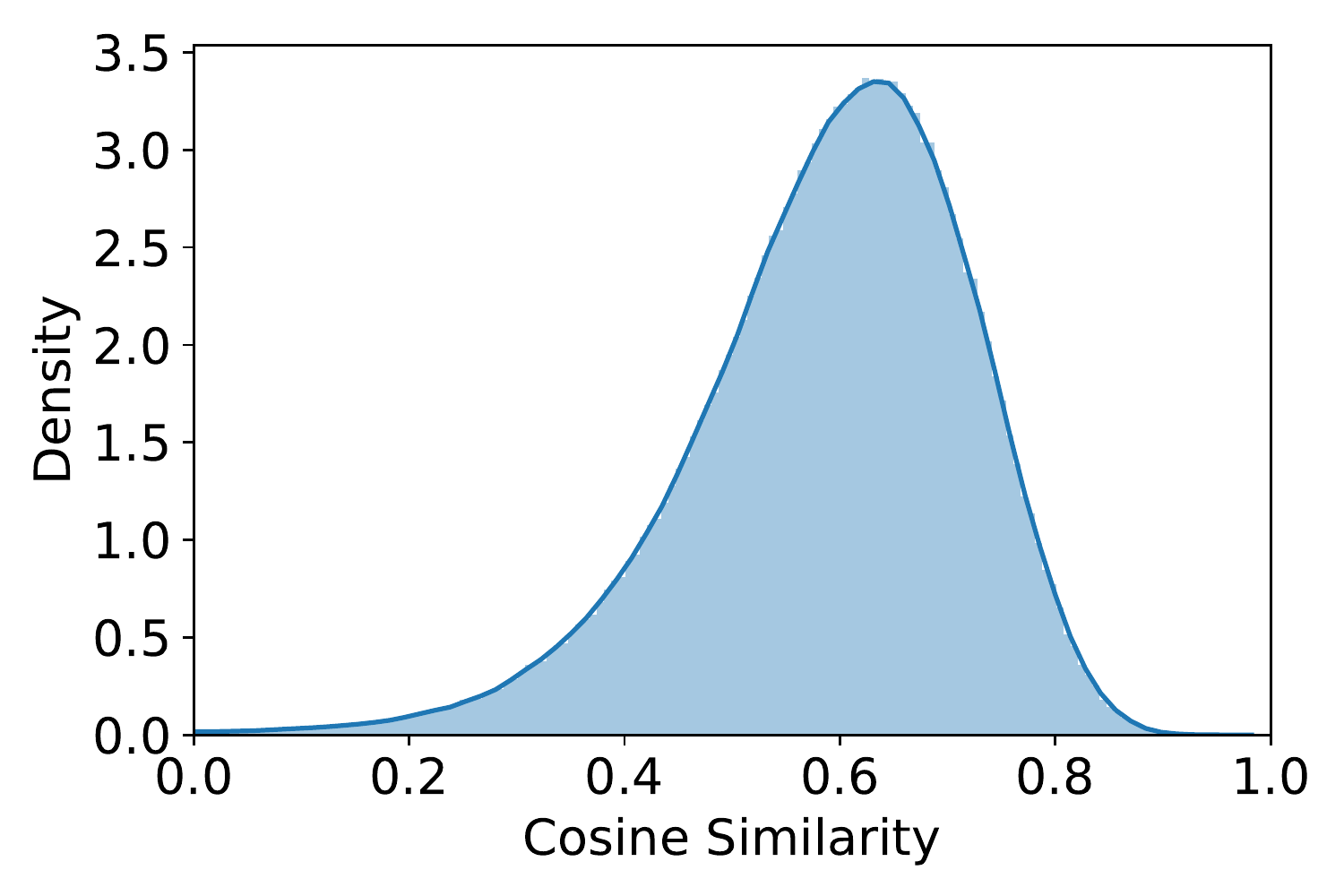}
	}
\caption{Distribution of news similarities.}\label{fig.dis}
\end{figure}

Following~\cite{wu2019nrms} we use the 300-dimensional GloVe~\cite{pennington2014glove} embedding to initialize the word embedding table.
Transformers in \textit{LeaDivRec} have 20 attention heads and each head output is 20-dimensional. 
The hyperparameter $\lambda$ is 20.
The optimization algorithm is Adam~\cite{kingma2014adam} and the learning rate is 1e-4.
Following~\cite{wu2020mind} we use AUC, MRR, nDCG@5 and nDCG@10 as the metrics for recommendation accuracy, and following~\cite{chen2018fast} we use the intra-list average distance (ILAD) and intra-list minimal distance (ILMD) of the top 5 and 10 ranked news as metrics for recommendation diversity.
We randomly repeat each experiment 5 times and report the average scores.
To measure the semantic similarity between candidate news, we use the average of GloVe embeddings of words in them to obtain their semantic embeddings, which is a widely used sentence similarity evaluation method due to its simplicity.\footnote{We do not use the news embeddings learned by the models because they are not comparable across different methods.}
We use cosine similarities between these embeddings to compute the similarity scores.
The distributions of news similarities on the two datasets are shown in Fig.~\ref{fig.dis}.
We can see that both datasets have very similar news semantic similarity distributions, which means that news domain texts may have some common semantic patterns.

\begin{table*}[t]
 \caption{Recommendation accuracy and diversity of different methods on \textit{MIND}.} \label{table.performance} 
\begin{tabular}{l|cccc|cccc}
\Xhline{1.5pt}
\multicolumn{1}{c|}{\textbf{Method}} & \textbf{AUC} & \textbf{MRR} & \textbf{nDCG@5} & \textbf{nDCG@10} & \textbf{ILAD@5} & \textbf{ILAD@10} & \textbf{ILMD@5} & \textbf{ILMD@10}  \\ \hline
EBNR                                 & 0.6690       & 0.3279       & 0.3549          & 0.4117           & 0.1512          & 0.1634           & 0.1273          & 0.1208           \\
DKN                                  & 0.6657       & 0.3220       & 0.3512          & 0.4086           & 0.1473          & 0.1566           & 0.1201          & 0.1113           \\
NAML                                 & 0.6773       & 0.3293       & 0.3602          & 0.4166           & 0.1688          & 0.1744           & 0.1375          & 0.1259          \\
NPA                                  & 0.6771       & 0.3313       & 0.3612          & 0.4186           & 0.1699          & 0.1758           & 0.1403          & 0.1269           \\
LSTUR                                & 0.6805       & 0.3344       & 0.3633          & 0.4193           & 0.1730          & 0.1769           & 0.1422          & 0.1306           \\
NRMS                                 & 0.6812       & 0.3340       & 0.3645          & 0.4199           & 0.1726          & 0.1775           & 0.1392          & 0.1301            \\
FIM                                  & 0.6826       & 0.3356       & 0.3648          & 0.4219           & 0.1739          & 0.1781           & 0.1410          & 0.1298          \\ \hline
SCENE                                & 0.6028       & 0.2746       & 0.2982          & 0.3665           & 0.1844          & 0.1856           & 0.1484          & 0.1335          \\
MMR                                  & 0.6632       & 0.3295       & 0.3489          & 0.4058           & 0.2139          & 0.2057           & 0.1663          & 0.1466         \\
DPP                                  & 0.6648       & 0.3210       & 0.3503          & 0.4076           & 0.2124          & 0.2031           & 0.1650          & 0.1444      \\
PD-GAN                               & 0.6610       & 0.3187       & 0.3466          & 0.4036           & 0.2087          & 0.1989           & 0.1614          & 0.1416    \\ \hline
LeaDivRec                              & 0.6757       & 0.3302       & 0.3596          & 0.4173           & 0.2147          & 0.2096           & 0.1669          & 0.1487       
               \\\Xhline{1.5pt}
\end{tabular}
\end{table*}

\begin{table*}[t]
 \caption{Recommendation accuracy and diversity of different methods on \textit{PrivateNews}.} \label{table.performance2} 
\begin{tabular}{l|cccc|cccc}
\Xhline{1.5pt}
\multicolumn{1}{c|}{\textbf{Method}} & \textbf{AUC} & \textbf{MRR} & \textbf{nDCG@5} & \textbf{nDCG@10} & \textbf{ILAD@5} & \textbf{ILAD@10} & \textbf{ILMD@5} & \textbf{ILMD@10} \\ \hline
EBNR                                 & 0.6295       & 0.3518       & 0.3846          & 0.4454           & 0.1443          & 0.1460           & 0.0873          & 0.0768        \\
DKN                                  & 0.6236       & 0.3469       & 0.3811          & 0.4397           & 0.1322          & 0.1347           & 0.0781          & 0.0673       \\
NAML                                 & 0.6370       & 0.3604       & 0.3926          & 0.4517           & 0.1500          & 0.1561           & 0.0955          & 0.0819      \\
NPA                                  & 0.6388       & 0.3611       & 0.3943          & 0.4524           & 0.1479          & 0.1560           & 0.0963          & 0.0819       \\
LSTUR                                & 0.6442       & 0.3668       & 0.3994          & 0.4578           & 0.1562          & 0.1593           & 0.0982          & 0.0866    \\
NRMS                                 & 0.6423       & 0.3658       & 0.3982          & 0.4567           & 0.1558          & 0.1564           & 0.0952          & 0.0821      \\
FIM                                  & 0.6441       & 0.3672       & 0.3996          & 0.4585           & 0.1545          & 0.1594           & 0.0970          & 0.0838        \\ \hline
SCENE                                & 0.5883       & 0.3167       & 0.3523          & 0.4102           & 0.1616          & 0.1624           & 0.0978          & 0.0835       \\
MMR                                  & 0.6273       & 0.3496       & 0.3818          & 0.4429           & 0.1895          & 0.1883           & 0.1084          & 0.0972        \\
DPP                                  & 0.6254       & 0.3478       & 0.3801          & 0.4410           & 0.2006          & 0.1943           & 0.1124          & 0.0989    \\
PD-GAN                               & 0.6212       & 0.3445       & 0.3776          & 0.4377           & 0.1844          & 0.1838           & 0.1066          & 0.0959           \\ \hline
LeaDivRec                              & 0.6337       & 0.3571       & 0.3889          & 0.4488           & 0.2059          & 0.1983           & 0.1167          & 0.1024            \\
 \Xhline{1.5pt}
\end{tabular}
\end{table*}

\subsection{Main Results}

We first compare the recommendation accuracy and diversity of several methods.
The methods to be compared include:
\begin{itemize}
    \item \textit{EBNR}~\cite{okura2017embedding}, embedding-based news recommendation, which uses autoencoders for news modeling and GRU networks for user modeling;
    \item \textit{DKN}~\cite{wang2018dkn}, deep knowledge network for news recommendation, which uses a knowledge-aware CNN to model news and uses candidate-aware attention to model user interest; 
    \item \textit{NAML}~\cite{wu2019}, an attentive multi-view learning based news recommendation method; 
    \item \textit{NPA}~\cite{wu2019npa}, using user ID embeddings as attention queries in personalized attention networks to model news and users; 
    \item \textit{LSTUR}~\cite{an2019neural}, using GRU to model short-term user interest and user ID embeddings to model long-term user interest; 
    \item \textit{NRMS}~\cite{wu2019nrms}, using multi-head self-attention networks for news and user modeling; 
    \item \textit{FIM}~\cite{wang2020fine}, a fine-grained interest matching method that uses 3-D CNN for user modeling;
\item  \textit{SCENE}~\cite{li2011scene}, reranking candidate news in the same topic based on their popularity;
\item  \textit{MMR}~\cite{carbonell1998use}, a reranking method that greedily selects candidate items based on a weighted combination of relevance and diversity;
\item  \textit{DPP}~\cite{chen2018fast}, a fast MAP inference method for determinantal point process to improve diversity via reranking;
\item  \textit{PD-GAN}~\cite{wu2019pd}, an adversarial learning-based method that uses a discriminator to classify whether the ranking list is given by the recommendation model (reranked by DPP) or generated by randomly sampling a user's interacted items in different categories.
\end{itemize}
For fair comparison, all these methods use news titles to model news content.
In addition, for reranking baselines, we apply them to the recommendation results generated by the best-performed non-diversity-aware methods.
The results on the \textit{MIND} and \textit{PrivateNews} datasets are shown in Tables~\ref{table.performance} and~\ref{table.performance2}, respectively.
From the results, we find the top recommendation results given by news recommendation methods without diversity consideration (e.g., \textit{NRMS} and \textit{FIM}) usually have low diversity.
This is because these methods only aim to optimize recommendation accuracy and candidate news are ranked solely based on their relevance to user interest, which will lead to low diversity among top ranked news.
In addition, we find that although the \textit{SCENE} reranking method can improve diversity, it has a huge sacrifice on recommendation accuracy, which yields a suboptimal accuracy-diversity tradeoff.
We also find that \textit{MMR} and \textit{DPP} have similar performance on the \textit{MIND} dataset, while \textit{DPP} slightly performs better than \textit{MMR} on the \textit{PrivateNews} dataset.
It may be because \textit{DPP} is stronger in optimizing the  tradeoff between accuracy and diversity.
However, we find that \textit{PD-GAN} does not outperform \textit{DPP}, which may be because users' click behaviors are sparse and clicked news may only cover a small number of categories.
Moreover, \textit{LeaDivRec} outperforms all other compared methods in terms of the tradeoff between recommendation accuracy and diversity, and the differences between \textit{LeaDivRec} and other baseline methods are significant ($p<0.01$ in two-sided t-test).
It shows the effectiveness of learning  diversity-aware recommendation model in an end-to-end manner.

\begin{figure*}[!t]
	\centering
	\subfigure[\textit{MIND}.]{
	\includegraphics[width=0.46\linewidth]{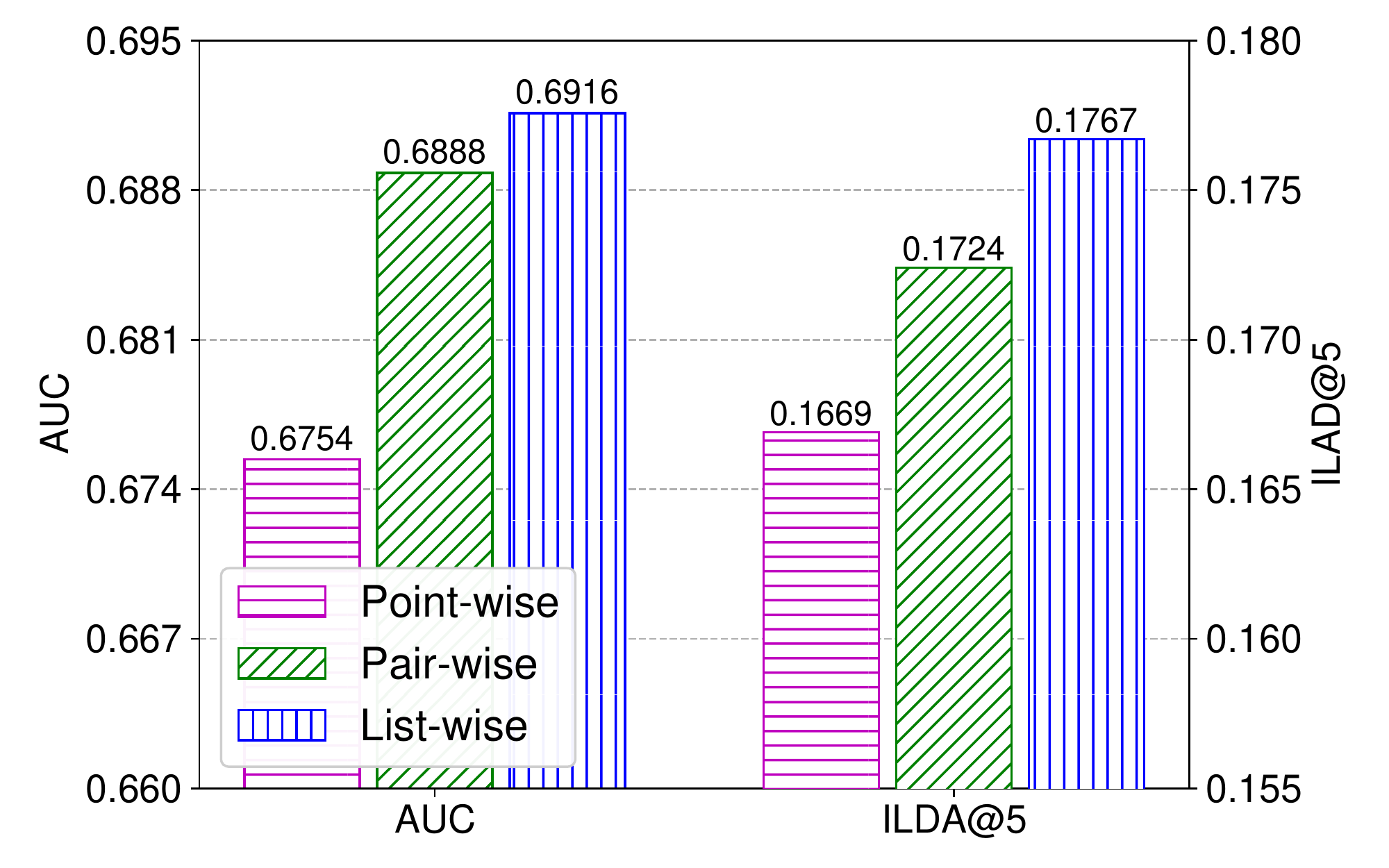}
	}
		\subfigure[\textit{PrivateNews}.]{
	\includegraphics[width=0.46\linewidth]{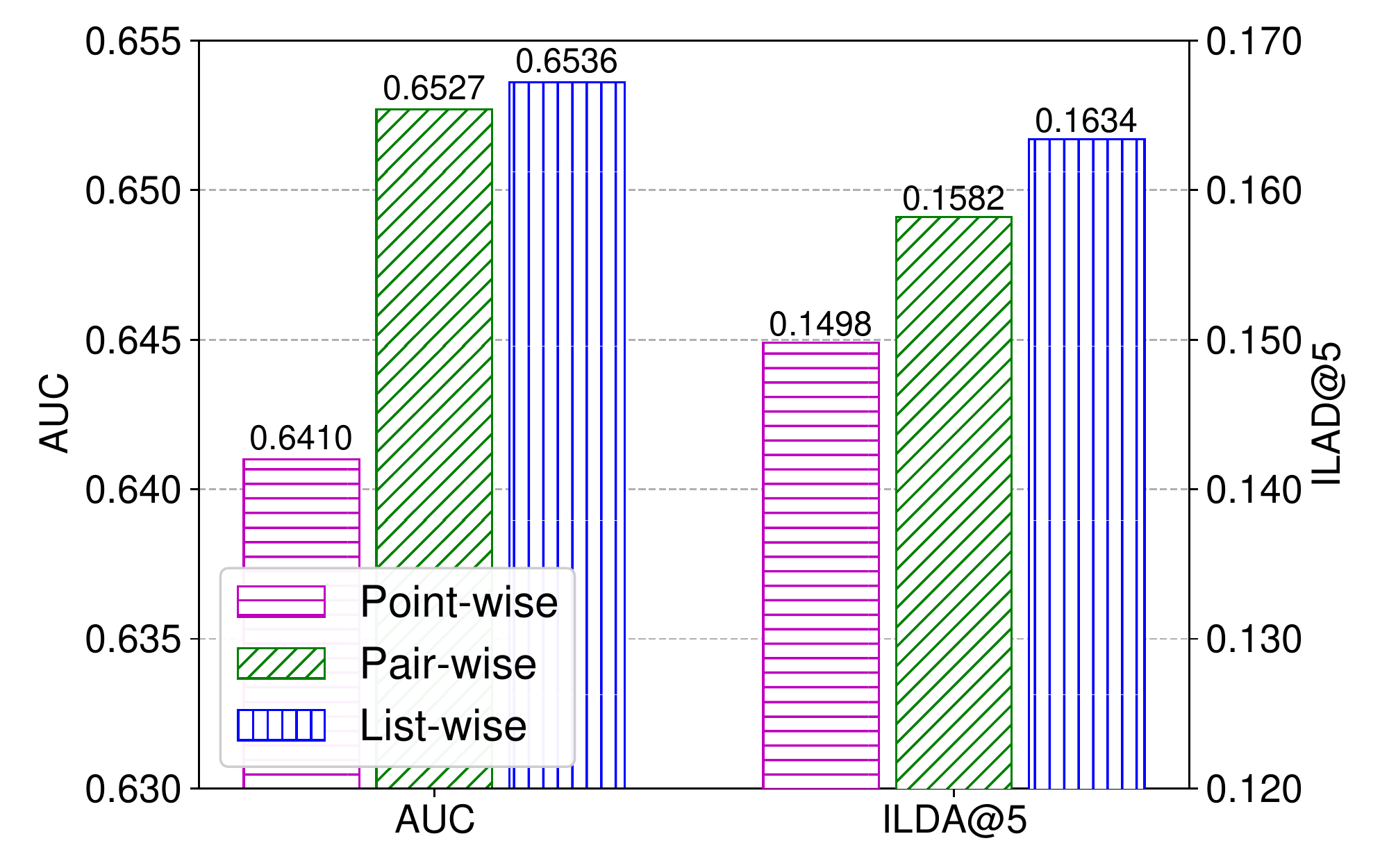}
	}
\caption{Effectiveness of list-wise ranking model.}\label{fig.rank}
\end{figure*}

\begin{figure*}[!t]
	\centering
	\subfigure[\textit{MIND}.]{
	\includegraphics[width=0.46\linewidth]{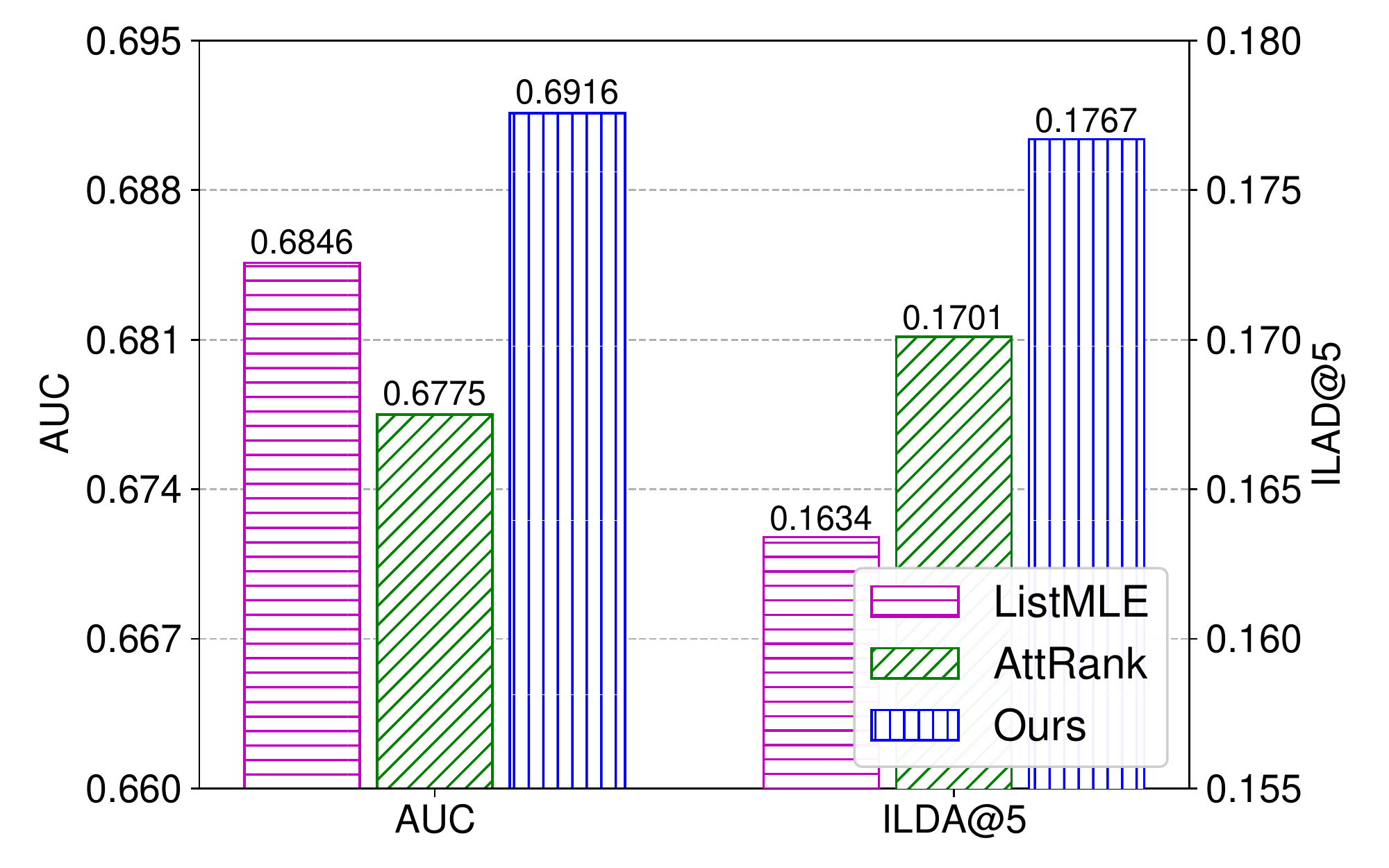}
	}
		\subfigure[\textit{PrivateNews}.]{
	\includegraphics[width=0.46\linewidth]{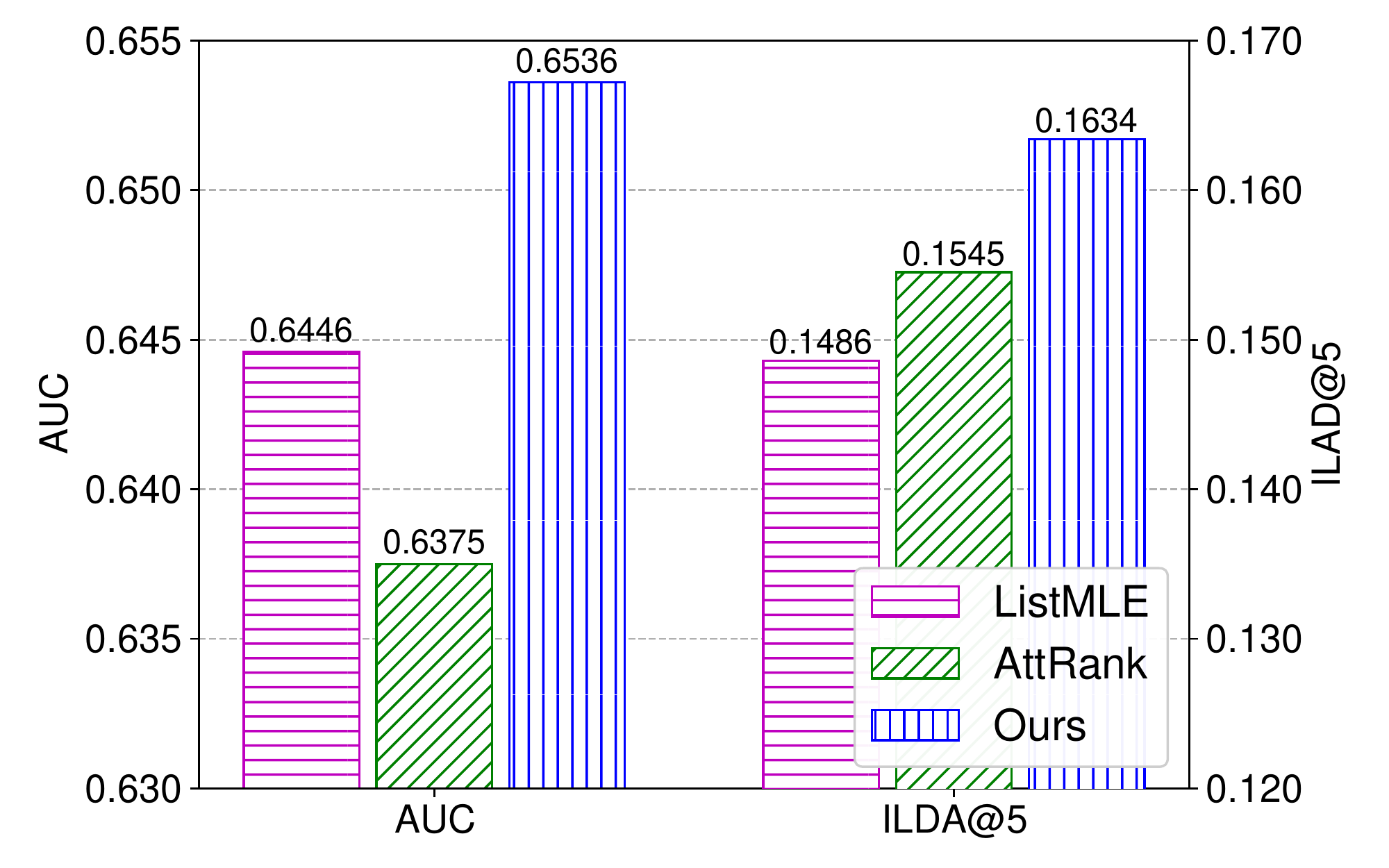}
	}
\caption{Comparison between our proposed list-wise recommendation model training method and several existing list-wise learning-to-rank methods.}\label{fig.rank2}
\end{figure*}

\subsection{Effectiveness of List-wise Ranking Model}

Next, we verify the effectiveness of list-wise candidate ranking in news recommendation.
Note that the diversity regularization loss is deactivated in this part.
We compare our proposed list-wise model training method with two widely used techniques in news recommendation model training.
The first one is regarding click prediction as a binary classification task (denoted as point-wise), and ranking candidate news based on the click scores.
The second one is training the recommendation model via BPR~\cite{rendle2009bpr} loss (pair-wise), and ranking candidate news based on click scores.
For fair comparison, the same news and user models are used.
The results on the two datasets are shown in Fig.~\ref{fig.rank}.
We find that the point-wise training method performs the worst.
This is because it handles different news in the candidate list independently, which is not optimal in training discriminative recommendation models.
In addition, the list-wise model outperforms the pair-wise model in terms of both accuracy and diversity.
This is because although the pair-wise method can exploit the relatedness between different candidate news in the training stage, it still ranks candidate news independently in the test stage.
Thus, it is also suboptimal in generating accurate and diverse recommendation lists.
These results show that list-wise ranking models have greater potentials in making both accurate and diversity-aware recommendations.

We then compare the list-wise recommendation model training method in our approach with two widely used list-wise learning-to-rank methods, i.e., \textit{ListMLE}~\cite{xia2008listwise} and \textit{AttRank}~\cite{ai2018learning}.
We also remove the diversity regularization loss here.
The results are shown in Fig.~\ref{fig.rank2}.
We find that our method outperforms both compared baseline methods.
This is because different from the search scenarios that  documents have different relevance ranks, in news recommendation there are only click/non-click signals, which limits the ability of \textit{ListMLE}.
In addition, since the click and non-click samples are extremely imbalanced, it is difficult for the \textit{AttRank} method to accurately recognize important candidates.
Different from these methods, our approach can contrast positive samples to negative ones, and can model the importance of impressions by adding the loss associated with positive samples within the same impression.
Thus, our approach is more suitable for list-wise news recommendation model training.

\begin{figure*}[!t]
	\centering
	\subfigure[\textit{MIND}.]{
	\includegraphics[width=0.46\linewidth]{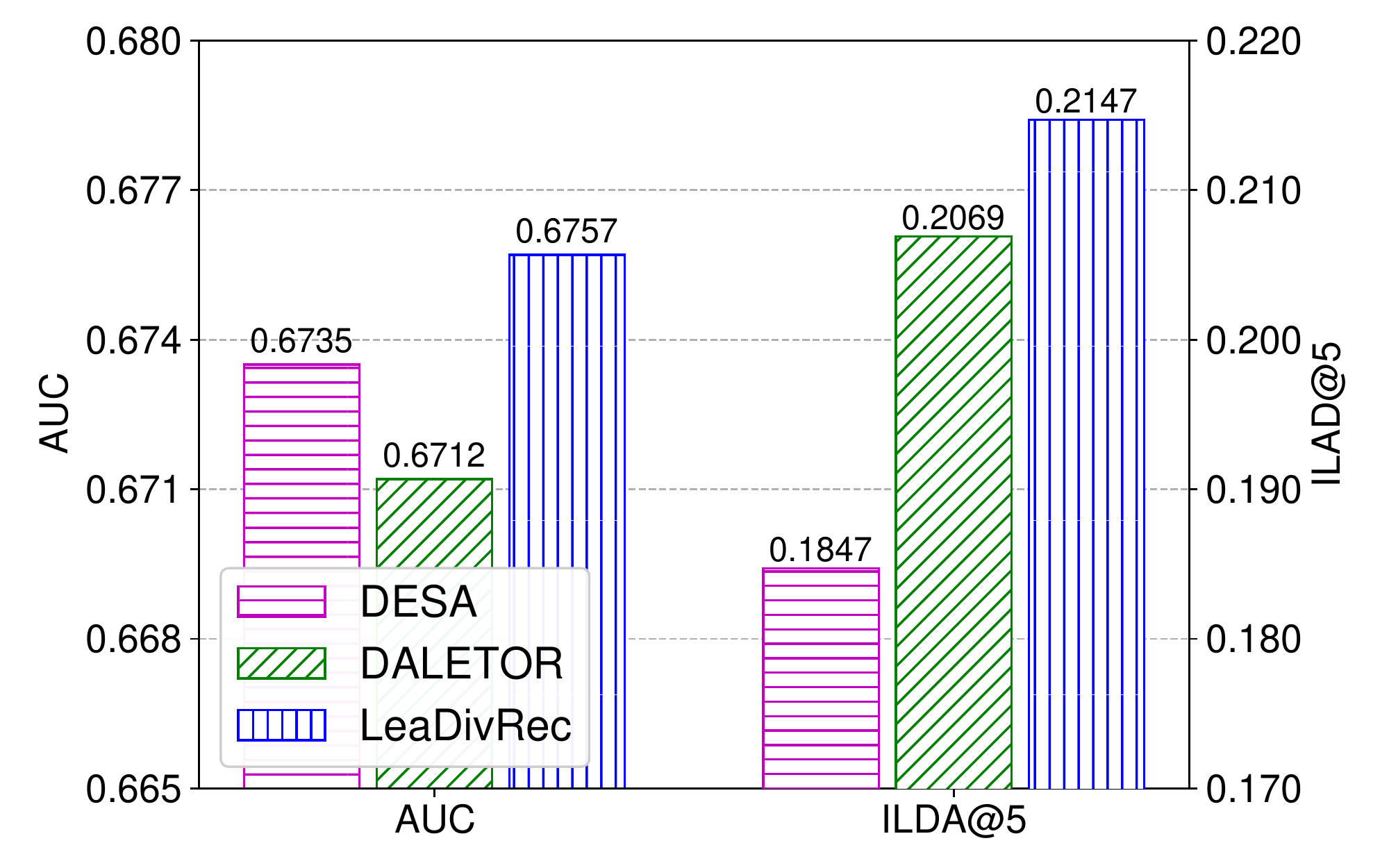}
	}
		\subfigure[\textit{PrivateNews}.]{
	\includegraphics[width=0.46\linewidth]{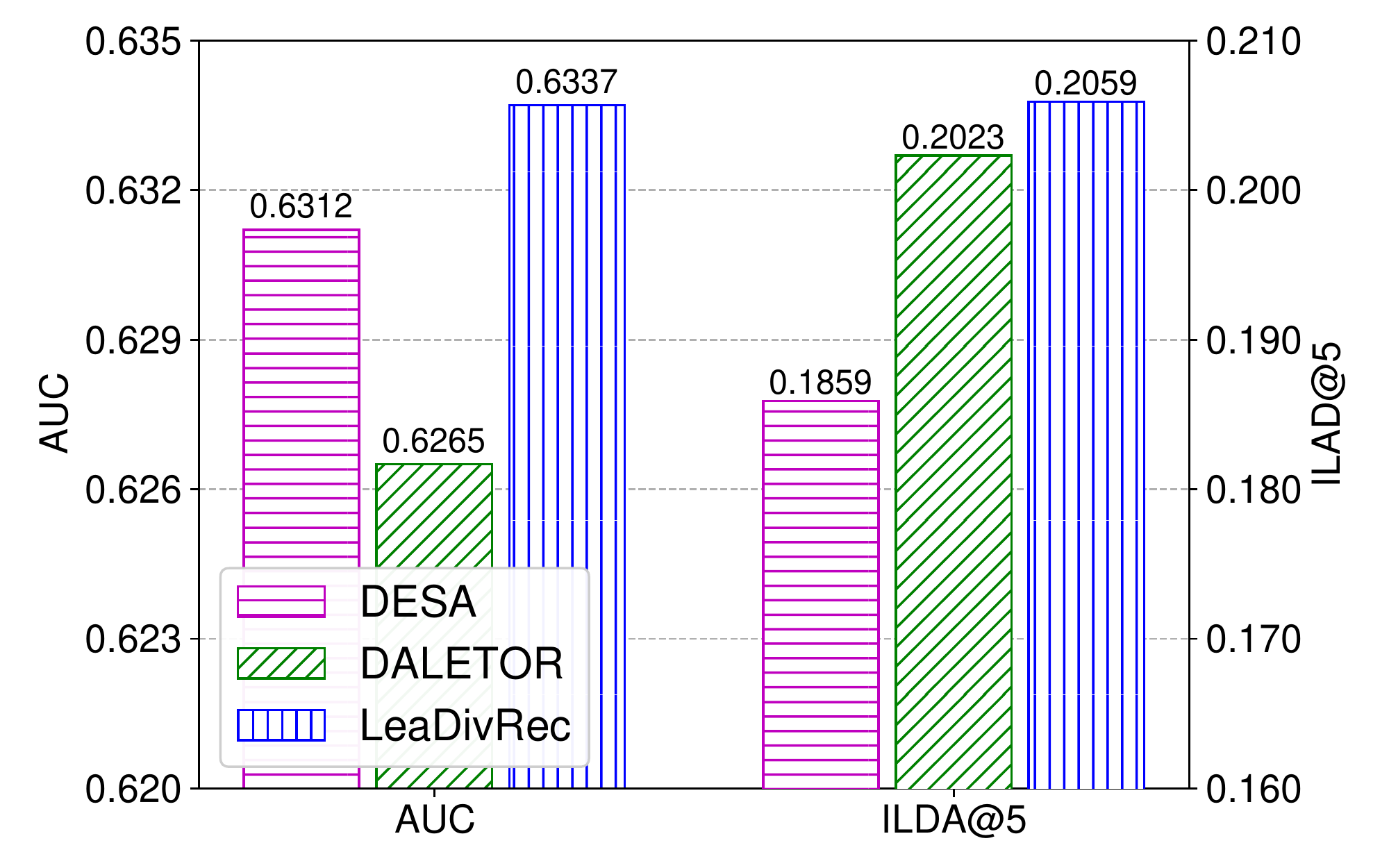}
	}
\caption{Comparison between \textit{DiverseRec} and several learnable search diversifying methods.}\label{fig.rank3}
\end{figure*}

\begin{figure*}[!t]
	\centering
	\subfigure[\textit{MIND}.]{
	\includegraphics[width=0.46\linewidth]{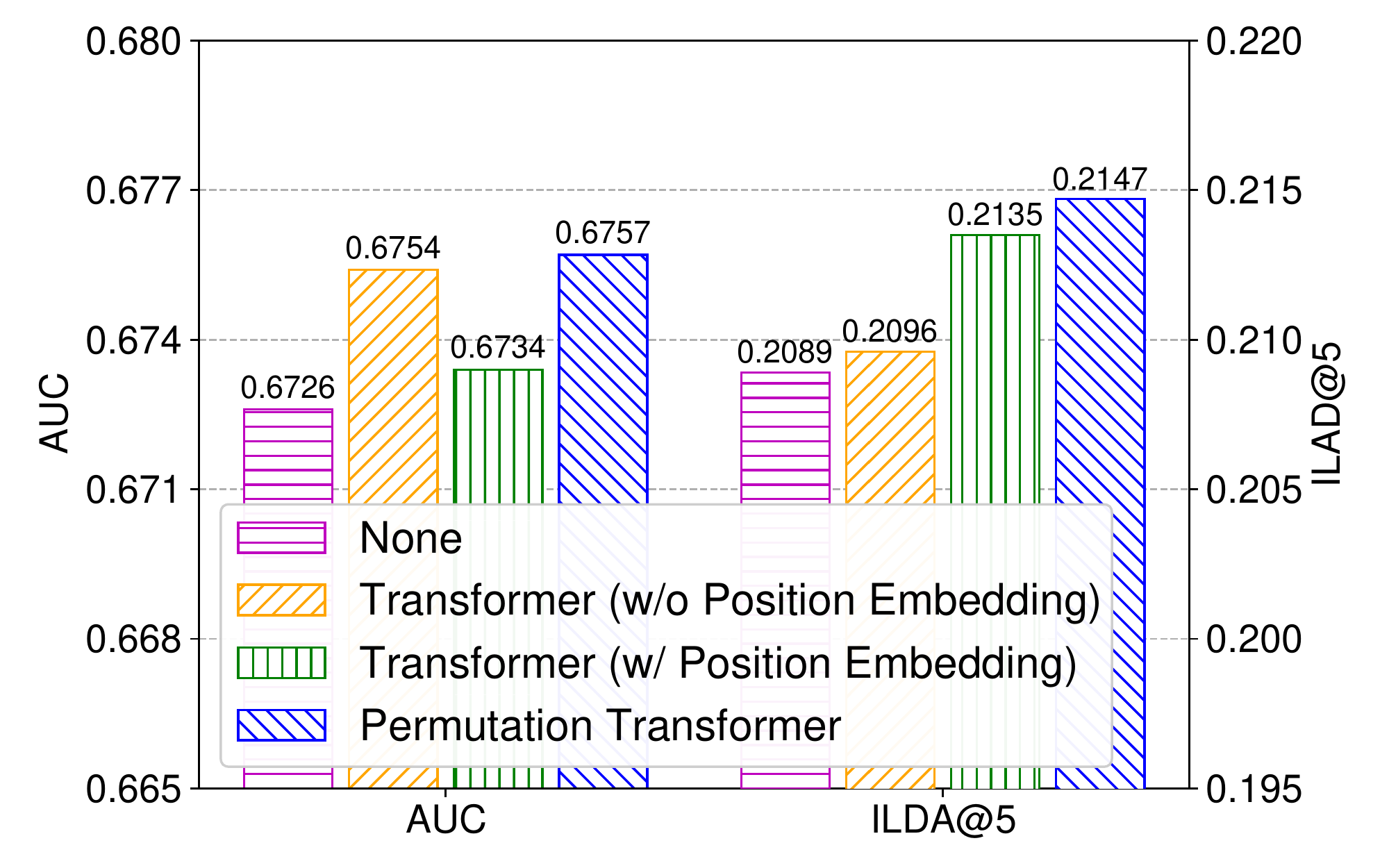}
	}
		\subfigure[\textit{PrivateNews}.]{
	\includegraphics[width=0.46\linewidth]{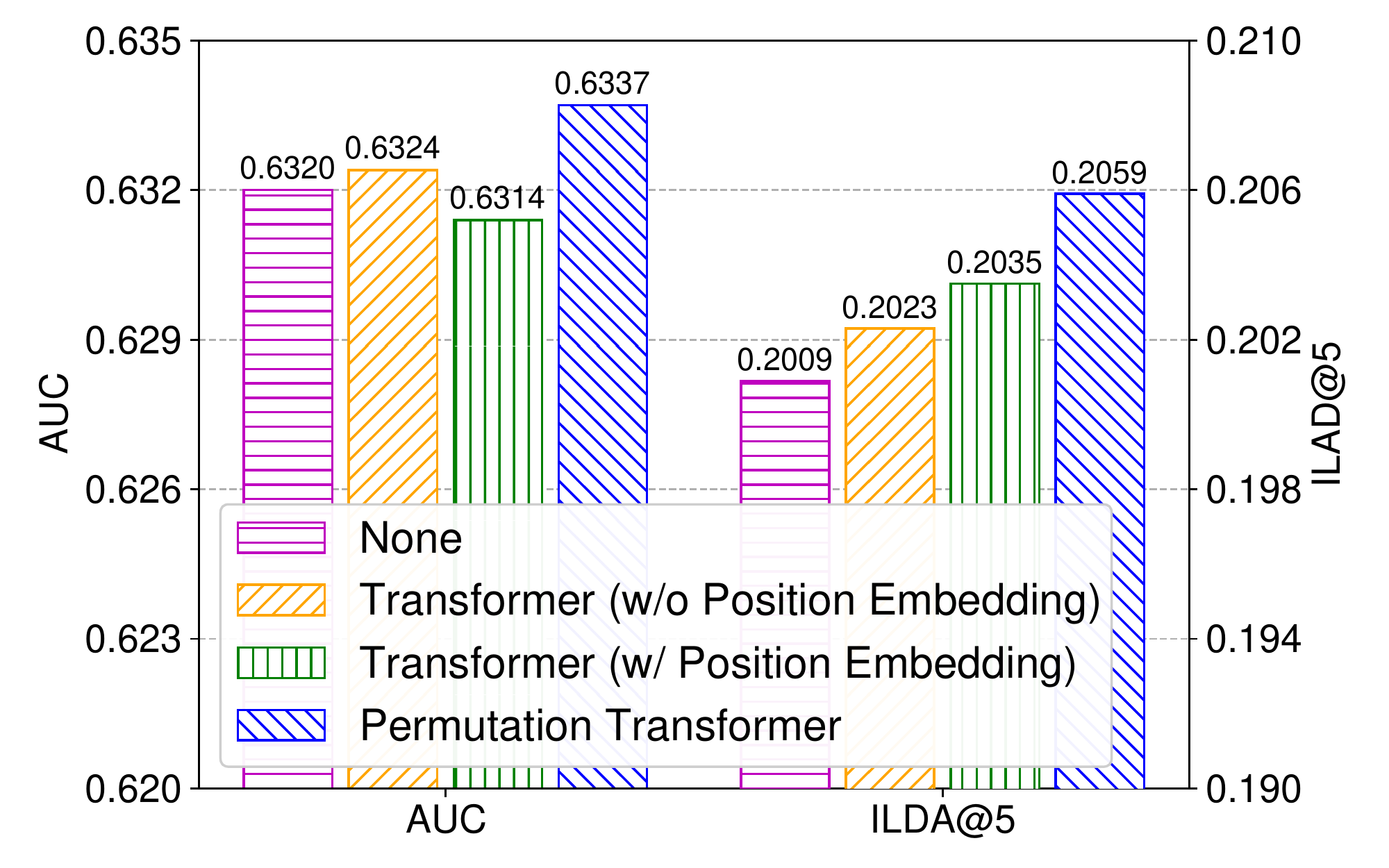}
	}
\caption{Effectiveness of permutation Transformer.}\label{fig.tf}
\end{figure*}

\subsection{Comparison with Search Diversifying Methods}

Furthermore, we compare \textit{LeaDivRec} with several list diversifying methods in the search ranking field, including \textit{DESA}~\cite{qin2020diversifying} and \textit{DALETOR}~\cite{yan2021diversification}.
The results are  shown in Fig.~\ref{fig.rank3}.
We find that the \textit{DESA} method has a satisfactory accuracy, it cannot effectively promote diversity.
This is because the self-attention mechanism used by \textit{DESA} cannot learn diverse representations for very similar candidate news, which leads to a limited diversity improvement.
Another compared method \textit{DALETOR} can better improve diversity than \textit{DESA}, but its accuracy sacrifice is much larger.
This is because it mainly considers the subtopic information of candidates to achieve diversification rather than their fine-grained semantics.
Our \textit{LeaDivRec} approach can improve recommendation diversity more effectively with less accuracy loss than these baselines, which further verify its effectiveness in balancing the accuracy and diversity of news recommendation.

\subsection{Effectiveness of Permutation Transformer}

Afterward, we conduct experiments to verify the effectiveness of our proposed permutation Transformer in the candidate list model.
We compare it with three variants, including (a) None, processing candidate news independently; (b) Transformer (w/o position embedding), using Transformers to model the relatedness between candidate news but does not incorporate position embeddings to keep it order-agnostic;
(c) Transformer (w/ position embedding), using Transformers with position embeddings to distinguish different candidate news.
The results are shown in Fig.~\ref{fig.tf}.
We can see that it is suboptimal to model different candidate news independently.
This is because different news in the candidate list may have some relatedness, which is important for making the recommendation decisions.
In addition, we find it is interesting that although using position embeddings can slightly improve recommendation diversity, it has some sacrifice on recommendation accuracy.
This is because the input candidate list is unordered and it may not be suitable to simply diversify candidate news representations by incorporating position embeddings.
Different from them, our proposed permutation Transformer has better performance in terms of both accuracy and diversity.
This is mainly because our approach can learn diverse representations for similar news and meanwhile may be less sensitive to input orders.

\begin{figure*}[!t]
	\centering
	\subfigure[\textit{MIND}.]{
	\includegraphics[width=0.46\linewidth]{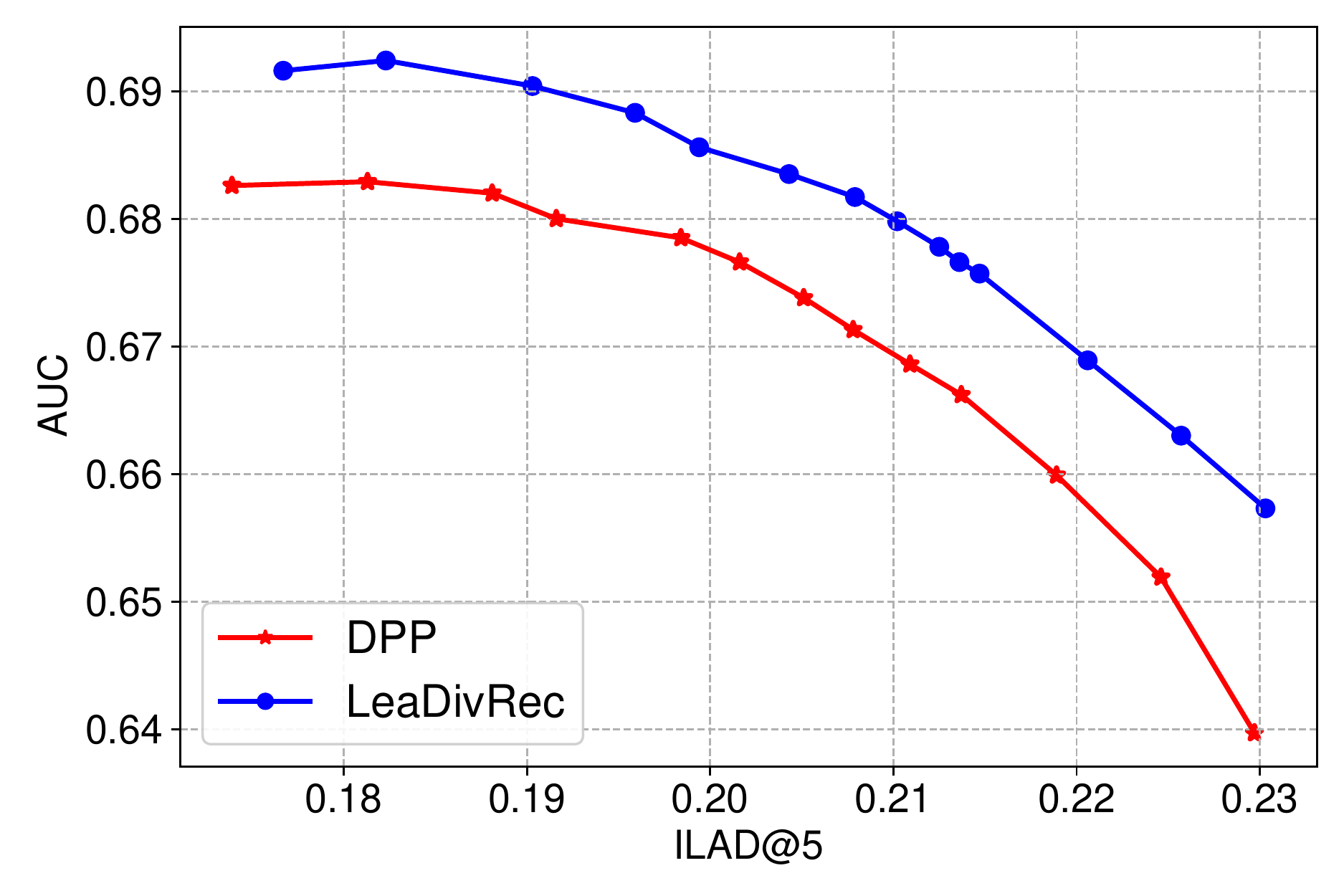}
	}
		\subfigure[\textit{PrivateNews}.]{
	\includegraphics[width=0.46\linewidth]{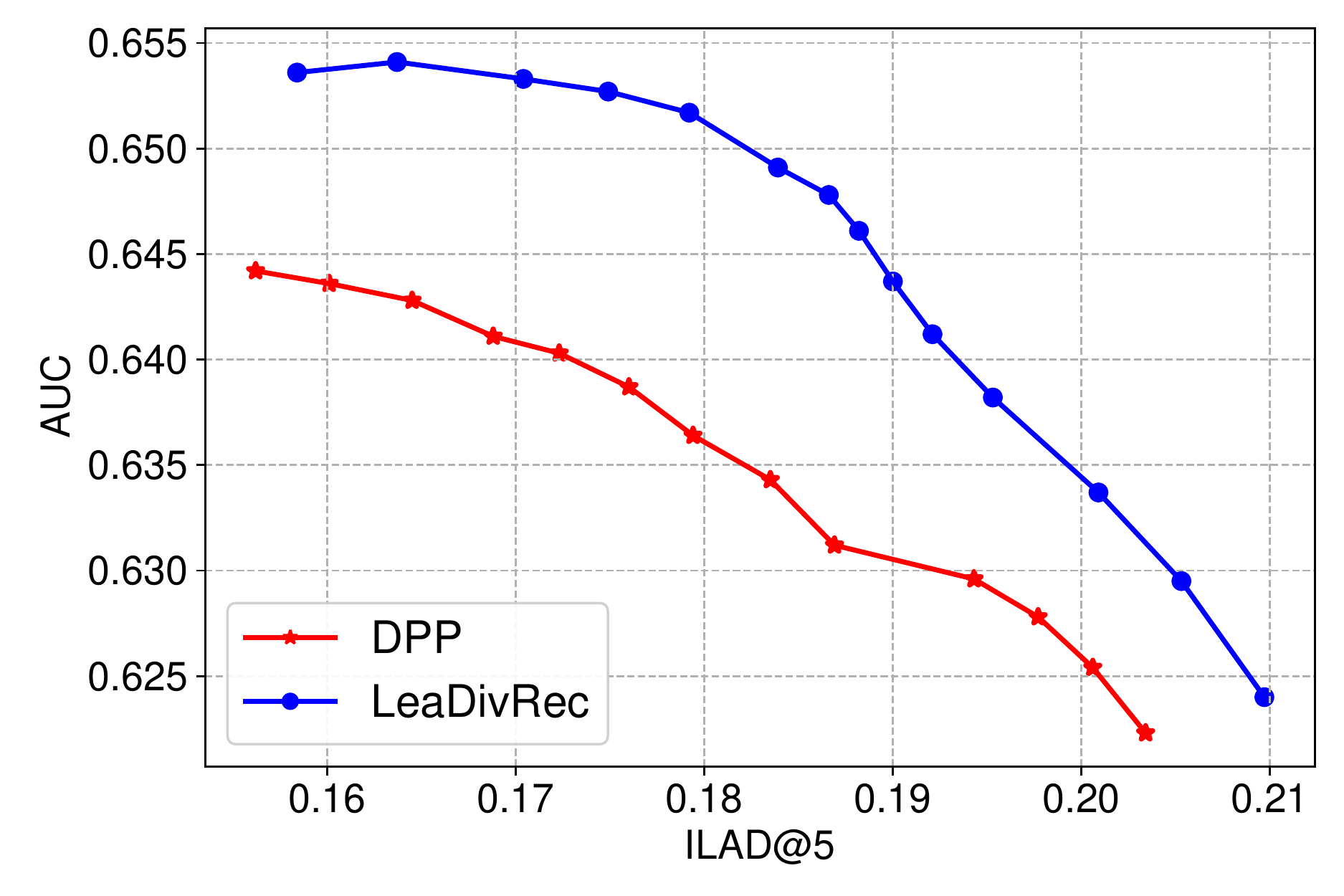}
	}
\caption{Tradeoff between recommendation accuracy and diversity.}\label{fig.acc}
\end{figure*}

\begin{figure*}[!t]
	\centering 
	\includegraphics[width=0.85\textwidth]{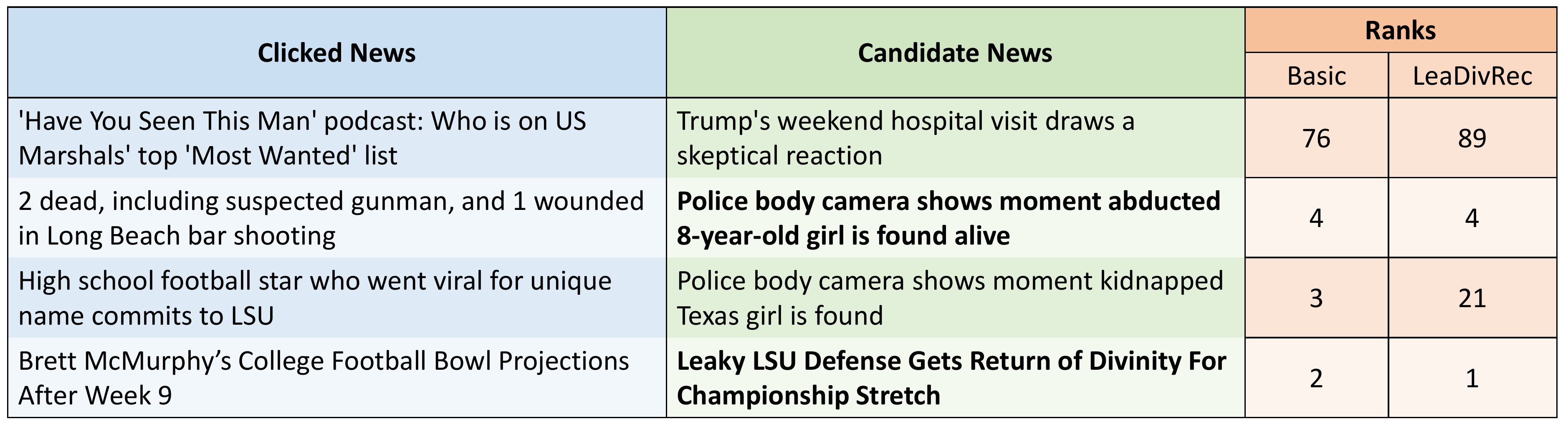}
\caption{The historical clicked news from a randomly selected user, as well as the candidate news and their ranks given by different methods. Clicked candidate news are in bold.}\label{fig.case}
\end{figure*}

\subsection{Tradeoff Between Recommendation Accuracy and Diversity}

We then analyze the tradeoff between recommendation accuracy and diversity of \textit{LeaDivRec} and \textit{DPP} (\textit{DPP} is applied to the recommendation results generated by \textit{FIM} on \textit{MIND} and \textit{LSTUR} on \textit{PrivateNews}).
In \textit{DPP} the value of $\theta$ is selected from $\{1.0, 0.98, ..., 0.9, 0.85,\\ 0.80, 0.75\}$\footnote{We do not use 0.99 due to the numerical stability issue.} and the value of $\lambda$ in \textit{LeaDivRec} is chosen from $\{0, 1, ..., 10,\\ 15, 20, 25\}$.\footnote{Lower $\theta$ values in \textit{DPP} mean stronger preferences with recommendation diversity.} 
The results are shown in Fig.~\ref{fig.acc}.
We find that \textit{DPP} achieves a good tradeoff between accuracy and diversity when $\theta=0.91$ on \textit{MIND} and $\theta=0.85$ on \textit{PrivateNews} (the performance drop becomes large when $\theta$ is larger).
In addition, \textit{LeaDivRec} achieves a good tradeoff when $\lambda=10$ on \textit{MIND} and $\lambda=20$ on \textit{PrivateNews}.
Moreover, we can see that \textit{LeaDivRec} can achieve higher recommendation accuracy than \textit{DPP} under the same diversity requirement, which shows that \textit{LeaDivRec} is more effective in balancing the accuracy and diversity of recommendation results.

\subsection{Case Study}

Finally, we present some case studies by comparing the ranking results given by different methods.
Fig.~\ref{fig.case} shows the historical clicked news and  candidate news of a user and the ranks given by the basic \textit{NRMS} model  and \textit{LeaDivRec}.
We can see that the first candidate news is irrelevant to user interests, while the rest three candidate news are related to the user interests in news indicated by clicked news such as sports and police related news.
However, the second and third candidate news are very similar and they mention the same event.
The basic model assigns these news neighboring ranks, which leads to less diverse recommendation results.
By contrast, \textit{LeaDivRec} can give the two similar news very different rankings.
In addition, the clicked one that has more detailed information gains higher ranks, which shows that our approach can provide both diverse and accurate recommendation results.
It implies the potentials of end-to-end models in diversity-aware recommendation.

%% file: data/conclusion.tex
\section{Conclusion}\label{sec:Conclusion}
In this paper, we propose an end-to-end diversity-aware news recommendation approach named \textit{LeaDivRec}, which can balance recommendation accuracy and diversity with a fully learnable model.
Different from existing point-wise and pair-wise news recommendation methods, in \textit{LeaDivRec} we propose a novel  list-wise news ranking model to better exploit the relatedness between candidate news within a candidate list.
More specifically, we propose a permutation Transformer to fully capture the relations among candidate news and meanwhile generate diverse representations for similar news to improve recommendation diversity.
In addition, we propose an effective list-wise model training method with diversity-aware regularization to learn the recommendation model given different diversity intensity preferences.
Extensive experiments on two real-world datasets validate that our approach can achieve better tradeoffs between recommendation accuracy and diversity than many baseline methods.